\documentclass[aip,apl,10pt,twocolumn,showpacs,preprintnumbers,amsmath,amssymb]{revtex4-1}

\usepackage{graphicx}
\usepackage{epstopdf}
\usepackage{dcolumn}
\usepackage{bm}

\usepackage{color}

\begin{document}

\title{Increasing the active surface of titanium islands on graphene by nitrogen sputtering}

\author{T.~Mashoff}
\author{D.~Convertino} 
\author{V.~Miseikis}
\author{C.~Coletti}
\affiliation{Center for Nanotechnology Innovation @ NEST, Istituto
Italiano di Tecnologia, Piazza San Silvestro 12, 56127 Pisa Italy}

\author{V.~Piazza}
\affiliation{Center for Nanotechnology Innovation @ NEST, Istituto
Italiano di Tecnologia, Piazza San Silvestro 12, 56127 Pisa Italy}

\author{V.~Tozzini}
\affiliation{NEST, Istituto Nanoscienze -- CNR and Scuola Normale
Superiore, Piazza San Silvestro 12, 56127 Pisa, Italy}

\author{F.~Beltram}
\affiliation{Center for Nanotechnology Innovation @ NEST, Istituto
Italiano di Tecnologia, Piazza San Silvestro 12, 56127 Pisa Italy}
\affiliation{NEST, Istituto Nanoscienze -- CNR and Scuola Normale
Superiore, Piazza San Silvestro 12, 56127 Pisa, Italy}

\author{S.~Heun}
\email{stefan.heun@nano.cnr.it}
\affiliation{NEST, Istituto Nanoscienze -- CNR and Scuola Normale
Superiore, Piazza San Silvestro 12, 56127 Pisa, Italy}

\date{\today}

\begin{abstract}
Titanium-island formation on graphene as a function of defect density is investigated. When depositing titanium on pristine graphene, titanium atoms cluster and form islands with an average diameter of about 10\,nm and an average height of a few atomic layers. We show that if defects are introduced in the graphene by ion bombardment, the mobility of the deposited titanium atoms is reduced and the average diameter of the islands decreases to 5\,nm with monoatomic height. This results in an optimized coverage for hydrogen storage applications since the actual titanium surface available per unit graphene area is significantly increased. 
\end{abstract}

%\pacs{...}
\maketitle

%\section{Motivation}
Hydrogen is one of the most promising energy carriers, particularly since its only combustion waste product is water.\cite{Schlapbach2001} Its practical exploitation, however, is hindered by several technical problems. Among these one of the most challenging is storage.\cite{Schlapbach2001,B802882F} In this respect, graphene recently attracted attention as a storage material owing to its chemical stability, low weight, and favorable physical-chemical properties for hydrogen adsorption.\cite{Elias30012009} Importantly it was shown that this storage capacity can be further increased by surface functionalization of graphene.\cite{PhysRevB.87.014102} Titanium was indicated as one of the most promising candidates for such functionalization.\cite{PhysRevB.77.085405,doi:10.1021/jp100230c} Theoretical calculations showed gravimetric densities of up to 7.8\,wt.\%.\cite{PhysRevB.77.085405} These estimates were based on the assumption of isolated titanium atoms positioned at the center of the graphene hexagons. Unfortunately titanium forms relatively large islands when deposited on a graphene surface.\cite{Mashoff:013903} Compared to individual atoms, islands present less binding sites per atom for hydrogen.\cite{Sun:14582} Indeed, as Ti islands grow larger, more and more atoms are in the bulk configuration and are expected not to contribute to the net hydrogen-storage capacity of the system, with a resulting significantly smaller hydrogen uptake. A possible way to achieve smaller islands or even individual titanium atoms on the graphene surface would be to reduce their mobility on the graphene surface, so that they would not cluster after deposition. Although this could in principle be done by cooling down the sample, such an approach is of no practical interest since upon the first annealing cycle island coalescence would occur and lead to an irreversible change in island morphology.

Here we present a different approach based on the controlled introduction of defects in the graphene layer. Defects change the electronic structure and can pin the titanium atoms to the defect sites themselves. Several calculations predicted a strong binding of the titanium atoms to defects in a graphene sheet.\cite{PhysRevB.87.014102,APL92:013106,Kim2009} Due to an increased charge transfer, the binding energy of the hydrogen molecules may be slightly lowered in the case of titanium on graphene with defects, but this is not expected to influence the stability at room temperature.\cite{PhysRevB.87.014102,APL92:013106,Kim2009}

%\section{Samples}
We used monolayer graphene grown on 4H--SiC(0001) as a substrate. It was obtained by annealing atomically-flat 4H--SiC(0001) samples for several minutes in argon atmosphere\cite{Starke2012} of 780\,mbar at about 1700\,K in a resistively heated cold-wall reactor (BM, Aixtron). Graphene quality and the actual number of layers were verified by atomic force microscopy and Raman spectroscopy.

%\section{Instrumental (UHV)}
All sample preparation and measurements were carried out in a two-chamber ultra high vacuum (UHV) system with a base pressure below $1\times10^{-10}$\,mbar. For preparation and analysis, the system is equipped with H$_2$ supply, sputter gun, heating/cooling stage (approx.~100\,K to 1300\,K), Ti-evaporator, and quadrupole mass spectrometer.  Characterization of the clean and processed surfaces was performed with a variable-temperature scanning tunneling microscope (STM). Details about the microscope can be found elsewhere.\cite{Goler2013249} After introducing graphene samples into the UHV system, they were annealed at 900\,K for several hours to remove water and other adsorbates. This was done by direct current heating of the substrate to ensure a homogeneous temperature. The latter was measured by a type K thermocouple at the position of the sample and cross-calibrated by an optical pyrometer.

%\section{Sputtering}
Defects in the graphene film were produced by molecular-nitrogen sputtering. To this end samples were positioned in the beam focus of the sputter gun, and a nitrogen pressure of $1.5\times 10^{-8}$\,mbar was applied to the preparation chamber through a needle valve. Sputter energies between 50\,eV and 300\,eV as well as sputter times from 30\,s to 8\,min were used to produce different defect patterns. Ion current was monitored by means of an ampermeter connected to the sample. The size and distribution of the resulting defects were analyzed by high-resolution STM imaging. 

\begin{figure}[tbhp]
  \centerline{\includegraphics[width=\columnwidth]{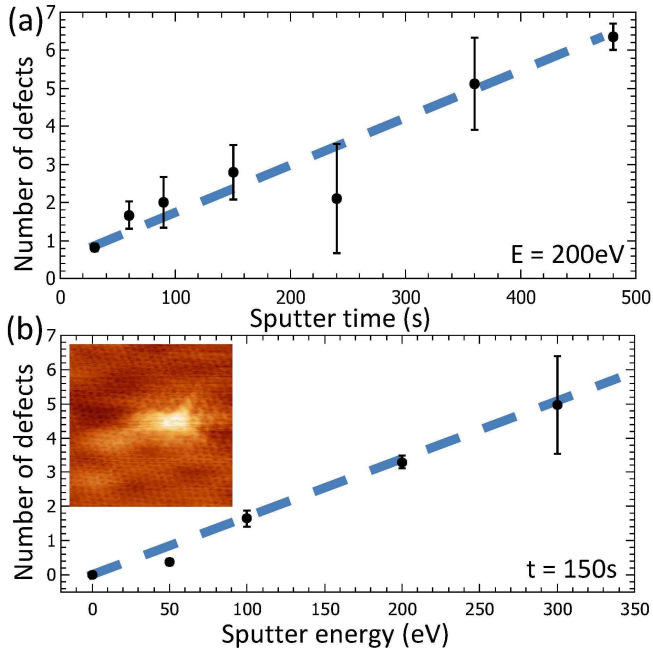}}
  \caption{(Color online) Number of defects per 100\,nm$^2$ induced in graphene by N$_2$-sputtering depending on (a) sputter time and (b) sputter energy. In both cases, the number of defects increases approximately linearly. The inset to (b) shows a $5\times 5$\,nm$^2$ STM image of an individual defect (V=1\,V, I=0.8\,nA).}
  \label{sputter}
\end{figure}

%\section{Results (Sputtering)}
Different types of defects can be induced by nitrogen sputtering depending on the sputter energy: nitrogen bombardment can create both vacancies and carbon-atom substitutions.\cite{Zhao:5062,PhysRevB.83.115424,PhysRevB.86.165428,Kim:2129} The latter defect type is more likely at low energies around 50\,eV, while there is a growing probability for vacancy formation with increasing energy. We did not observe any particular trend in the size or shape of the individual defects created by varying sputter time or energy. However, since similar hydrogen binding properties are predicted for titanium pinned at the two defect types,\cite{Kim2009,APL92:013106} we did not distinguish between the type of defect, but merely counted their total number. Additionally we measured the average size of the distortion in the electronic structure as seen by STM. 

As can be seen in Fig.~\ref{sputter}, the defect density increases linearly with (a) sputter time and (b) with sputter energy. The graphs show the counted number of defects per 100\,nm$^2$, averaged over several images. Error bars are the standard deviation of the average. The time--dependent measurements were taken at a constant sputter energy of E = 200\,eV while the energy--dependent measurements were taken at a constant sputter time of t = 150\,s. Annealing the surface for t = 10\,min at a temperature of T = 900\,K did not change the distribution of the defects. The inset of Fig.~\ref{sputter}(b) shows a $5\times 5$\,nm$^2$ STM image (V=1\,V, I=0.8\,nA) of a defect, which was created by sputtering at 100\,eV, and the atomically-resolved graphene surface.

\begin{figure}[tbhp]
  \centerline{\includegraphics[width=\columnwidth]{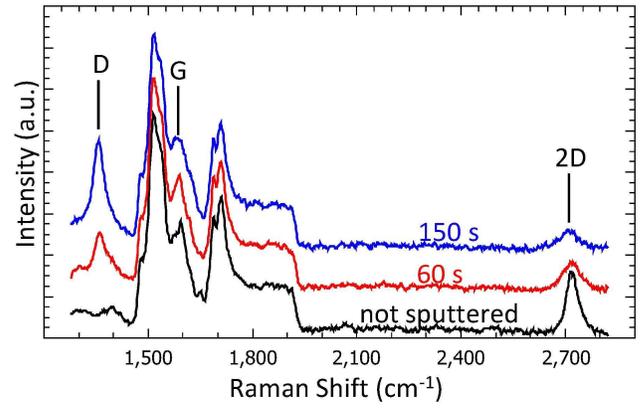}}
  \caption{(Color online) Raman spectra of the graphene samples. D, G, and 2D-peaks are marked. Other features originate from the substrate. The black curve shows the non-sputtered sample, the red curve shows a sample which was sputtered for 60\,s, with an ion energy of E = 200\,eV, and the blue curve corresponds to 150\,s of sputtering with identical parameters (offset for clarity). The height of the D-Peak increases with increasing number of defects, while the intensity of the 2D-Peak decreases.}
  \label{raman}
\end{figure}

Figure~\ref{raman} shows the change in the Raman spectra induced by sputtering (Laser wavelength $\lambda=532$\,nm). The D, G, and 2D-peaks are marked. The other features originate from the SiC substrate. The D-peak is located around 1360\,cm$^{-1}$, it originates from the breathing modes of the hexagonal rings and requires defects for its activation.\cite{Cancado:3190,Ferrari:187401} It is not present in pristine graphene and increases in intensity with increasing disorder. The 2D-peak (historically also known as G') is located at 2720\,cm$^{-1}$. It is the second order of the D-peak. Since it originates from a process where two phonons with opposite wavevectors ensure momentum conservation, no defects are required and thus it is also present in pristine graphene. Nevertheless, the process is influenced by the density of defects, and thus the intensity of the 2D-peak decreases for higher sputter rates, in good agreement with previous reports.\cite{Cancado:3190}

\begin{figure}[tbhp]
  \centerline{\includegraphics[width=\columnwidth]{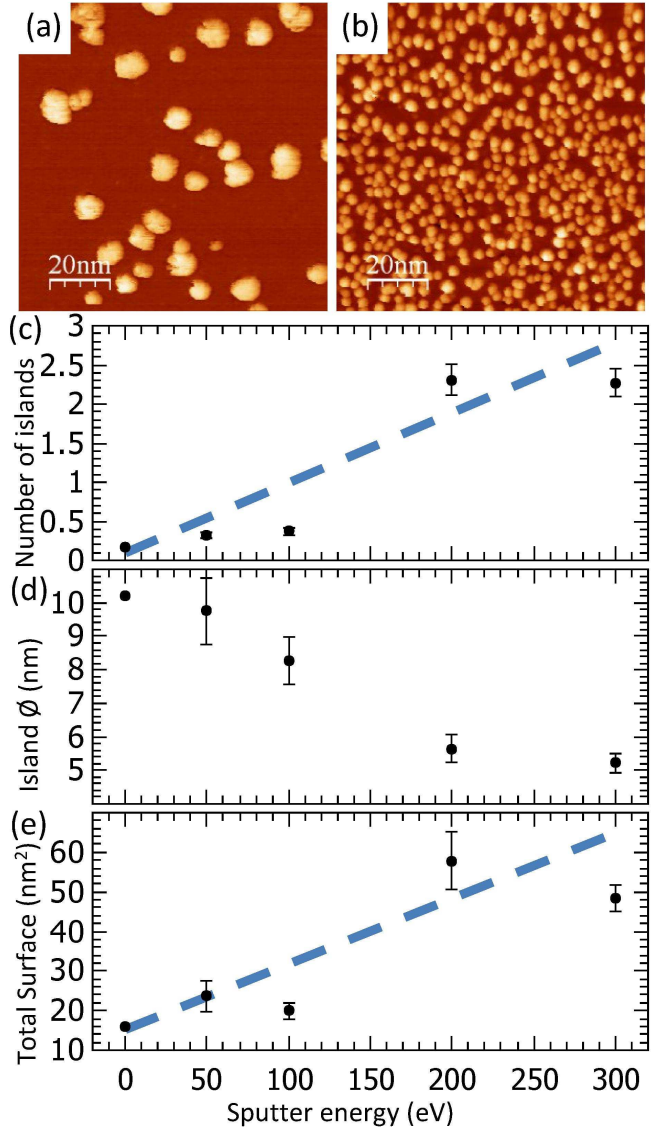}}
  \caption{(Color online) (a) $100\times 100$\,nm$^2$ STM image of the distribution of 0.55\,ML titanium on a pristine graphene surface. (b) Similar STM image for Ti-deposition after sputtering the sample at E = 300\,eV for 150\,s. The density of the islands is strongly increased with the number of induced defects, and their individual size decreases. (c) Number of islands per 100\,nm$^2$ for a sputter time of 150\,s and different sputter energies. (d) Average diameter of the Ti-islands for different sputter energies. (e) Evolution of the three--dimensional surface area of the Ti-islands normalized to a 100\,nm$^2$ sample region depending on the sputter energy with a constant sputter time of 150\,s. The total surface area of the islands approximately increases by a factor of $4$. $r_{Ti-G}$ defined in the text is obtained dividing the total surface area by 100\,nm$^2$.}
  \label{islands}
\end{figure}

%\section{Results (Islands)}
Following defect creation by sputtering, we deposited titanium onto the surface. The total amount of deposited titanium for all samples was 0.55\,ML (1\,ML = $1.32\times 10^{15}$\,atoms/cm$^{2}$), as calibrated by STM. The observed change in the distribution of titanium as a function of the different sputtering parameters is shown in Fig.~\ref{islands}. For small sputter energies up to approximately 100\,eV that yield a rather low defect density, we registered little change in island distribution. The number of islands increased very slowly and their average diameter decreased by less than 20\%. On the contrary, when we increased the sputter energy (and therefore the defect density) to 200\,eV and more, we observed a significant increase in the density of the titanium islands and a marked decrease in their size. We counted approximately 10 times more islands per unit area with respect to pristine graphene, and observed an average reduction in their diameter by more than a factor of two. 

Figure~\ref{islands}(a) shows a $100\times 100$\,nm$^2$ STM image of Ti-islands deposited on a pristine graphene surface. Relatively few islands are present, their average diameter exceed 10\,nm and their height is few ($2-3$) atomic layers. Sputtering the sample for 150\,s at an ion energy of E = 300\,eV before titanium deposition leads to a much higher density of islands as shown in Fig.~\ref{islands}(b). Here island diameters are around 5\,nm and heights are of one atomic layer only.

Figure~\ref{islands}(c) shows the increase in the number of titanium islands per 100\,nm$^2$ as a function of sputter energy with a constant sputter time of 150\,s. The corresponding size of the islands is shown in Fig.~\ref{islands}(d). The measured surface area of the titanium islands normalized to a sample region of 100\,nm$^2$ is plotted in Fig.~\ref{islands}(e). It shows an increase of the actual titanium surface by approximately a factor of 4 for an intensely sputtered sample with respect to Ti-deposition on a pristine graphene surface. It is not possible to see the defects in the graphene sheet underneath the titanium islands by STM, but we never saw any defects in the uncovered regions between the titanium islands. We conclude that the islands were indeed formed on top of the defects. Comparing Fig.~\ref{sputter}(b) and Fig.~\ref{islands}(c) leads to the observation that the number of induced defects is approximately twice as high as the number of titanium islands at any given sputter energy. This indicates that there is often more than one defect underneath an individual island. Importantly, titanium islands were stable at least up to T = 900\,K: annealing for 10\,min did not change the distribution as measured by STM.

\begin{figure}
  \centerline{\includegraphics[width=\columnwidth]{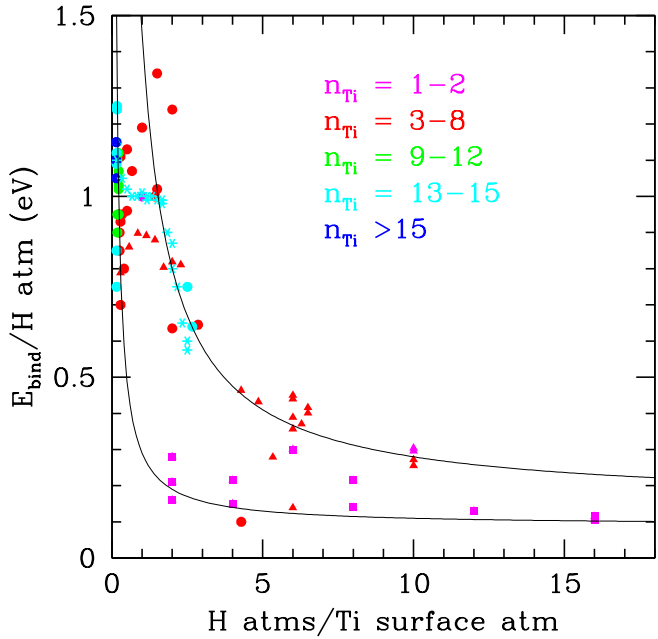}}
  \caption{A survey of the available theoretical studies (Density Functional Theory--based) on hydrogen adhesion on Ti clusters. The binding energy per H atom is reported as a function of the coverage (H atoms per surface Ti atom). Cluster size n$_{\rm Ti}$ (coded in the dots color as reported) ranges from 1 to 15 Ti atoms. Higher binding energy (0.5--1.5 eV/atm) denotes chemisorbtion, lower energy ($\sim 0.3$ eV/atm) denotes physisorption, however also intermediate or hybrid situations are present. The general trend indicates a decreases of the binding energy with increasing coverage, and a tendency towards physisorption for high coverage, which, in fact, can sustain up to 8 H$_2$ molecules per Ti atom. Conversely, only at most 2 H atoms per Ti atom can be chemically coordinated on a Ti surface atom. Intermediate situations correspond to configurations with part of the surface Ti atoms saturated with chemisorbed H and additional molecules physisorbed in the interstitial areas or non-saturated sites. The black lines are guides to the eye (represented by $A+B/x$ functions), intended to indicate the boundaries of the scatter of the data. Data are taken from Refs.~\onlinecite{tarak,shang,rojas,kumar,chu,liu}.}
  \label{theo}
\end{figure}

In summary, we investigated the dependence of titanium island distribution on the number of defects introduced in graphene. We showed that titanium atoms deposited after nitrogen sputtering are less mobile on the surface compared with pristine graphene and pin onto the defects. This leads to more and significantly smaller titanium islands, since it is no longer possible for the atoms to move large distances and agglomerate with other titanium atoms into large islands. This results in a larger surface available for hydrogen binding per unit graphene area by a factor $\simeq 4$, as shown in Fig.~\ref{islands}(e). Two effects contribute to this enhancement as the island size decreases: (i) the fraction $r$ of Ti atoms in the surface of the islands relative to the total island Ti--atom number increases, and (ii) the exposed Ti--surface to total graphene--surface ratio $r_{\rm Ti-G}$ increases (compare Figs.~\ref{islands}(a) and (b)). Both quantities can be directly measured. $r$ scales as $1/d$, the island diameter $d$ reported in  Fig.~\ref{islands}(d). However, small islands ($d<5$\,nm) are of monolayer height, therefore all atoms are surface atoms, i.e. $r = 1$. Larger islands ($d \sim 10$ nm) are $2-3$ layers high, so $r \sim 0.3 - 0.5$. Also $r_{\rm Ti-G}$ is directly measured, see Fig.~\ref{islands}(e).   

These quantities can be used to evaluate the Gravimetric Density (GD) of the system, i.e. the ratio of loaded-hydrogen mass over total system mass. In order to obtain GD, one also needs to know the average number
of loaded hydrogen molecules per Ti atom, $n_{\rm H_2}$. This was estimated as a function of island size in several theoretical studies,\cite{tarak,shang,rojas,kumar,chu,liu} summarized in Fig.~\ref{theo}. If one considers the spread of the plotted data and the characteristic cluster sizes relevant for the present study, one can safely estimate $n_{\rm H_2} = 1$ H$_2$ molecule per Ti atom. The general formula for GD can be written as
\begin{eqnarray}
{\rm GD} & = & \frac{M_{\rm H}}{M_{\rm Ti}+ M_{\rm G}+ M_{\rm H}} \nonumber \\
& = & \frac{n_{\rm H_2}m_{\rm H_2}r_{\rm Ti-G}}{ m_{\rm Ti} r_{\rm Ti-G} /r +   m_{\rm C} \sigma_{\rm G} / \sigma_{\rm Ti}   +n_{\rm H_2}m_{\rm H_2}r_{\rm Ti-G}} \nonumber
\end{eqnarray}
with $m_{\rm x}$ the atomic or molecular masses, and $\sigma_{\rm x}$  the surface particle density.  $\sigma_{\rm Ti}$ can be evaluated from the interatomic distance to be $13.2$ atoms/nm$^2$, which leads to $\sigma_{\rm G}/\sigma_{\rm Ti}\simeq 2.88$. All the involved quantities can be measured. 

Therefore, in the small islands regime $r=1$ and $r_{\rm Ti-G} = 0.55$ leading to GD $\simeq 1.8\%$. This number could be increased up to $2.4\%$ by increasing $r_{\rm Ti-G}$ to 1, corresponding to almost complete Ti coverage of the graphene sheet with small islands. In addition, there is a marked trend of larger hydrogen uptake for smaller islands. If we consider the few--Ti--atom limit, we expect $n_{\rm H_2} = 4$ for all the regimes examined in the literature. By inserting this value in the  previous equation we obtain GD=6.8\%, close to the theoretical limit of about 7.8\%.\cite{PhysRevB.77.085405} On the other hand, in the large islands regime $r_{\rm Ti-G} \simeq 0.15 - 0.2$ (see Fig.~\ref{islands}(e)) and $r\simeq 0.3-0.5$, leading to GD $\simeq 0.5 - 0.75\%$ for $n_{\rm H_2} = 1$.

In conclusion, we have demonstrated that a controlled introduction of defects in graphene reduces the size of Ti islands on such surfaces, while at the same time their number is increased, which overall results in an increased surface area of the Ti islands for a given amount of deposited Ti. We show that this increases the gravimetric hydrogen storage density from $0.5 - 0.75\%$ for pristine graphene to $2 - 2.5\%$ for these samples. Reducing the island size further, up to $7\%$ seem feasible.

%\begin{acknowledgments}
We acknowledge financial support from the CNR in the framework of the agreement on scientific collaboration between CNR and JSPS (Japan), joint project title 'High-mobility graphene monolayers for novel quantum devices', and from the Italian Ministry of Foreign Affairs, Direzione Generale per la Promozione del Sistema Paese. We also acknowledge funding from the European Union Seventh Framework Programme under grant agreement no. 604391 Graphene Flagship.
%\end{acknowledgments}

\bibliography{Torge-biblio}

\end{document}